\documentclass[amssymb, preprintnumbers, showpacs, showkeys, aps, prb, superscriptaddress, twocolumn, longbibliography]{revtex4-2}

\usepackage{color} 
\usepackage{hyperref}
\usepackage{slashed}
\usepackage{graphicx}
\usepackage{amsmath}
\usepackage{latexsym}
\usepackage{epstopdf}
\usepackage{amsmath}
\usepackage{enumitem}
\usepackage{amssymb,amsmath}
\usepackage{multirow}
\usepackage{mathtools}
\usepackage{bbm}
\usepackage{bm}
\usepackage{amsfonts}
\usepackage{amssymb}
\usepackage{calligra}
\usepackage{calrsfs}
\usepackage{makecell}
\usepackage[normalem]{ulem}
\usepackage[british,UKenglish,USenglish,english,american]{babel}
\usepackage{verbatim}
\usepackage{overpic}

\newcommand{\mean}[1]{\left\langle {#1} \right\rangle}

\begin{document}

\title{Helicity modulus in the bilayer XY model by the Monte Carlo worm algorithm}

\author{A. Masini}
\email{masini.alessandro94@gmail.com}
\affiliation{SuperComputing Applications and Innovation Department, CINECA, Via Magnanelli 6/3, Bologna, 40033, Italy}
\affiliation{Dipartimento di Fisica e Astronomia, Universit\`a di Firenze, I-50019, Sesto Fiorentino (FI), Italy}

\author{A. Cuccoli}
\email{alessandro.cuccoli@unifi.it}
\affiliation{Dipartimento di Fisica e Astronomia, Universit\`a di Firenze, I-50019, Sesto Fiorentino (FI), Italy}
\affiliation{INFN, Sezione di Firenze, I-50019, Sesto Fiorentino (FI), Italy}

\author{A. Rettori}
\email{angelo.rettori@unifi.it}
\affiliation{Dipartimento di Fisica e Astronomia, Universit\`a di Firenze, I-50019, Sesto Fiorentino (FI), Italy}
\affiliation{INFN, Sezione di Firenze, I-50019, Sesto Fiorentino (FI), Italy}


\author{A. Trombettoni}
\email{atrombettoni@units.it}
\affiliation{Dipartimento di Fisica, Universit\'a di Trieste, Strada Costiera 11, I-34151 Trieste, Italy}
\affiliation{SISSA and INFN Sezione di Trieste, Via Bonomea 265, I-34136 Trieste, Italy}
\affiliation{CNR-IOM DEMOCRITOS Simulation Center, Via Bonomea 265, I-34136 Trieste, Italy}

\author{F. Cinti}
\email{fabio.cinti@unifi.it}
\affiliation{Dipartimento di Fisica e Astronomia, Universit\`a di Firenze, I-50019, Sesto Fiorentino (FI), Italy}
\affiliation{INFN, Sezione di Firenze, I-50019, Sesto Fiorentino (FI), Italy}
\affiliation{Department of Physics, University of Johannesburg, P.O. Box 524, Auckland Park 2006, South Africa}

\date{\today}

\begin{abstract}
The behavior of the helicity modulus has been frequently employed to investigate the onset of the topological order characterizing the low-temperature phase of the two-dimensional XY-model. We here present how the analysis based on the use of this key quantity can be applied to the study of the properties of coupled layers. To this aim, we first discuss how to extend the popular worm algorithm to a layered sample, and in particular to the evaluation of the longitudinal helicity, that we introduce taking care of the fact that the virtual twist representing the elastic deformation one applies to properly define the helicity modulus can act on a single layer or on all of them. We then apply the method to investigate the bilayer XY-model, showing how the helicity modulus can be used to determine the phase diagram of the model as a function of temperature and inter-layer coupling strength.
\end{abstract}


\maketitle

\section{Introduction}\label{section1}


The two-dimensional (2D) XY model has been extensively used to understand, characterize, and produce qualitative and quantitative predictions for a plethora of phenomena typical 
of physical systems in two spatial dimensions at finite temperature \cite{kosterlitz2016} or in one spatial dimension at zero temperature \cite{giamarchi2004}. Indeed, this model features the celebrated Berezinskii-Kosterlitz-Thouless (BKT) transition \cite{berezinskii1972destruction,kos73,kosterlitz1974} and it has therefore been extensively used as the prototypical model to verify, analytically and numerically, the predictions and scalings of the BKT theory \cite{nobel2016,kosterlitz2016,kosterlitz2017}. Moreover, it allows to study in details the behavior of vortex excitations \cite{tobochnik1979}, the role of the vortex core \cite{jose1977,maccari2020}, the universal jump of the superfluid stiffness and of the superfluid density \cite{nel77,prokofev2000}, and the phenomenology of the unbinding of vortex-antivortex pairs \cite{kos73,simanek94}. The XY model 
has been studied via a variety of approaches, from the Monte Carlo 
\cite{tobochnik1979,gupta1988,prokofev2010worm,komura2012,svistunov15} 
and other numerical techniques \cite{honda1997,kuhner2000,yu2013,chatelain2014} to renormalization group \cite{jose1977,simanek94,grater1995,nagy2009,jakubczyk2014,defenu2017}, self-consistent harmonic approximation \cite{pokrovski1974magnetic,pires1996,cuccoli2000,giachetti2021} and field theory methods \cite{amit1980,nandori2001,mudry2014}. A further important aspect is that the 2D XY model, and the related Villain model \cite{Villain1975}, can be mapped (under certain well-discussed approximations) onto the 2D Coulomb gas \cite{minnhagen1987} and the (1+1) sine-Gordon model \cite{malard2013}. These mappings considerably helped to reconstruct common properties of 2D systems. This also created a well-known and widely used link with the theory of integrable field theories in low dimensions \cite{mussardo2010}.

An important advantage of the XY model relies in its versatility. 
Indeed, one can add, for instance, several ingredients relevant for applications and study their effect. Among many examples the 2D XY model has been studied in presence of magnetic fields 
\cite{fishman1987,benfatto2007}, of inhomogeneous couplings, 
\cite{giuliano2023} and of long-range couplings \cite{giachetti2021bis,giachetti2022,giachetti2023,defenu2023,sbierski2024} (see \cite{gulacsi1998,martinoli2000,drouintouchette2022} for more references). One can add a quantum kinetic term, giving rise to the so-called quantum phase model \cite{fazio2001} and there studying 
its quantum phase transitions at zero temperature 
\cite{simanek94,sac11}, possibly adding frustration charges \cite{bruder1992,grignani2000} or non-local interactions \cite{fishman1988,vanotterlo1995,capriotti2003}. 


Making reference to the BKT critical behavior of the XY model, and the underlying
role played by vortex excitations, revealed to be
useful also to describe the properties of real magnetic
compounds in different setups, ranging from the
thermodynamic behavior of quantum antiferromagnets
in presence of weakly easy-plane exchange anisotropies \cite{2003CRVV, 2003CRTVV}
or effective anisotropy induced by a uniform
field \cite{2003CRVVa,KohamaJA-VM-DMCM11}, to the behavior of the alternating 
magnetization induced by impurities in antiferromagnets \cite{EggertSAA07,2009CV,2010CV}.

Turning back to the pure XY model, by coupling different 2D XY systems one can explore the crossover from 2D to three dimensions (3D) \cite{lawrence1971theory,schneider2000,iazzi2012,rancon2017,peppler2018,guo2020}, which has been the subject of intense research due to its relevance, e.g., for the modelization of layered superconductors \cite{schneider2000,tinkham2004} and for adsorbed He films \cite{kotsubo1983,gallet1989}. There are two main ways to implements this crossover.

The first is to take a finite numbers of 2D XY models and couple them. Denoting by $M$ the number of layers and by $L$ the size (i.e., with a number of sites $L \times L$) of each 2D system, one keeps fixed $M$ and then increase $L$ to explore the thermodynamical limit. The case $M=2$ refers to the bilayer XY model \cite{parga1980,jiang1996,PhysRevLett.123.100601,song2022}.

In the second approach, one takes a 3D XY model on a $L \times L \times L$ with two different kinds of couplings: $J$ is the coupling in the 2D layers and $J_\perp$ the one among layers \cite{shenoy1995,schneider2000}. In this second approach, $J$ and $J_\perp$ are taken fixed and $L$ is increased, and one can study in such way the connection between 2D vortices and 3D vortex lines \cite{williams1987,shenoy1989}. In the first approach, to which one refers as the {\it multilayer} XY model, one expects and finds that the critical temperature goes from the the 2D BKT critical temperature of the 2D XY model (for $M=1$) to the 3D XY critical temperature (for $M \to \infty$, where the limit $L\to \infty$ is performed before). Also in the second approach one interpolates between the 3D and the 2D critical 
temperatures, namely, moving the ratio $\lambda=J_\perp/J$ from $\lambda=1$ (corresponding to the 3D XY model) to $\lambda=0$. This second approach refers to the so-called {\it layered} (or anisotropic) 3D XY model. 

The main difference between the multi-layer model and the layered model is in their critical properties. Indeed, for finite $M$ one expects that in the multilayer model the critical point (whose properties are determined in the $L\to \infty$ model) is BKT like, while in the layered model the critical point is expected to be in the 3D XY universality class for any finite value of the ratio $\lambda$. 

Here, we focus on the multilayer XY case and develop the {\it worm algorithm} (WA) for a generic value of $M$. We then apply it to the specific case of a bilayer, i.e., $M=2$. In the framework of the generalized elasticity theory for the XY model \cite{chaikin2000principles}, we examine the properties of the free energy at critical points. 
To this aim it is especially relevant the behavior of the helicity modulus, a key quantity which has been routinely employed to characterize the properties of XY models \cite{vh1981,olsson1991,harada1997} and more generally of superfluid systems \cite{fisher1973,chaikin2000principles}. We are particularly interested in examining our results in the light of the recent predictions about the phase diagram of the bilayer XY model obtained in Ref. \cite{PhysRevLett.123.100601}, where three phases were discussed. The first is the high-temperature phase, where the (two-point-in-layer and four-point-inter-layer) 
correlation functions decay exponentially, in contrast to the low-temperature phase, where both the 
correlation functions decays as a power law featuring quasi-off-diagonal long-range order \cite{RevModPhys.34.694}, as typical of the BKT low-temperature phase \cite{berezinskii1972destruction,kos73,kosterlitz1974}. 
In addition, a third, intermediate regime, where the two-point correlation function decays exponentially but the four-point correlation decays as power law, also emerges from the results in \cite{PhysRevLett.123.100601}. Both the numerical and analytical results there reported support the conclusion that the intermediate phase appears for any finite coupling between different layers. 
However, when considering continuous spins residing on a two-dimensional-like geometry, solely fitting the two-point correlation function proves to lead to results heavily dependent on the system size, especially close to the critical temperature. It is therefore mandatory to provide additional explorations using refined methods. Moreover, the analysis in \cite{PhysRevLett.123.100601} does not characterize the superfluid properties of the intermediate phase and the response of the system in the different regimes to a perturbation such as the torsion that defines the helicity modulus. 
In the framework of finite-size scaling analysis, the study of the helicity estimator then offers a quantitative robust approach to probe the BKT character of the transitions we are interested to look into. It also give insights on the superfluid properties of the system.
Based on \cite{PhysRevLett.123.100601}, one may expect that the presence of the intermediate region in the phase diagram affects the behavior of the helicity modulus and the question arises as to how the response of the system differs in the different regimes and especially in the intermediate paired phase. In the following we are going to explore these issues by the WA.


This paper is organized as follows: In Section~\ref{section2} we introduce the model Hamiltonian, whereas Section~\ref{section1bis} presents the worm algorithm for the investigated geometry, as well as the necessary estimators for the physical relevant quantities. Our findings for the helicity modulus are presented in Section~\ref{section4}, where we also discuss the connection with the results of Ref. \cite{PhysRevLett.123.100601}. Section~\ref{section5} is devoted 
to our conclusions.

\section{The multilayer model}\label{section2}

The present section introduces the model Hamiltonian we will employ to investigate a 
(coupled)  
multilayer system. As it is well-known, a straightforward manner to explore the critical properties of a system belonging to the BKT class is to consider a simple planar spin model \cite{kos73}, as such a simple model allows us to come up with general considerations about a large variety of experimental systems \cite{Weitering_2017,Taborek2020,Ren_2016}. 
Getting back to the problem discussed in Section~\ref{section1}, a planar model on a multilayer geometry with $M$ layers
can be described by the Hamiltonian:
\begin{multline}
\label{eq:ham}
H=-\sum_{m=1}^M \left[ J \sum_{\langle ij\rangle}  \cos \left(\theta_{i,m}-\theta_{j,m}\right) \right.\\
\left.+K\sum_{i=1}^N \cos \left(\theta_{i,m}-\theta_{i,m+1}\right) \right]\,, 
\end{multline}
$\theta_{i,m}$ representing the angle of a unitary and planar spin $\vec{s}_{i,m}$=$\left(\cos\theta_{i,m},\sin\theta_{i,m} \right)$ with respect to an arbitrary, irrelevant reference direction. 
The index $i$ labels the position of the spin on the $m$-th layer, with $m=1,\cdots,M$. For the sake of clarity we denote by the unit vectors $\hat{\mathbf{x}}$ and $\hat{\mathbf{y}}$ the main directions in the layers, whereas $\hat{\mathbf{z}}$ is the orientation perpendicular to the layers. The spins are arranged on a simple cubic lattice with a unitary lattice spacing, so that nearest-neighboring layers are coupled between them in the $z$ direction. 
For a generic value of $M>2$, periodic boundary condition 
are imposed along all the directions of the system. For $M=2$, 
referring to the bilayer case, when there are only two layers, we do not impose periodic boundary conditions on the $z$-direction. For simplicity, again in the $M=2$ case, we denote the angles of the rotators as
$$\theta_{i,m=1} \equiv \theta_i; \, \, \, \theta_{i,m=2}=\phi_i\,, $$
with the $\theta_i$'s for the first layer ($m=1$) 
and the  $\phi_i$'s for the second layer ($m=2$). 

In order to get a perfect ferromagnetic state at zero temperature, both $J$ and $K$ must be considered as positive. The coupling constant $J$ is referring to nearest neighbor in-layer spins, $\langle ij\rangle$. At the same time  $K$ describes the interactions between spins located on nearest neighboring slabs. In the limit $K\to0$, the system is simply made of uncorrelated layers, each of them showing the usual BKT behavior. In the opposite case, $K \to \infty$, the interaction term constrains spins on nearest-neighbor layers to have the same orientation, allowing us to describe the model in terms of an effective two dimensional Hamiltonian whose critical temperature will increase with $M$, being substantially proportional to $M$ when $K\gg J$. 
From now on we will express $K$ as well as the temperature (setting $k_B=1$) in units of $J$.

Interesting features have been recently highlighted by Monte Carlo simulations, and further supported by a renormalization group analysis, for $M=2$ and $0<K\lesssim2.5$~in Ref. \cite{PhysRevLett.123.100601}. 
In particular, the in-layer two-point correlations $\mean{\cos{(\theta_i-\theta_j)}}$ display the onset of an algebraic behavior at a lower temperature if compared to the out-of-layer four-point ones $\mean{\cos{(\theta_i+\phi_i-\theta_j-\phi_j)}}$. This brings to the conclusion of the existence of a quasi-long-range order for pairs of spins on different layers. Such a phase 
has been referred to as a topological BKT-\textit{paired phase} in \cite{PhysRevLett.123.100601}. 
The study of correlations has been of key importance to understand the physics of a pairing system marked by a continuous symmetry, and it 
is very useful to look into its features at and near the critical points. This
has been accomplished extending the Monte Carlo examination to observable such as, for instance, the generalized rigidity, also known as the helicity modulus \cite{chaikin2000principles,PhysRevB.83.174415}, that we are going to discuss in the next Section.

\section{Worm Algorithm for multilayer models}\label{section1bis}

To efficiently sample the properties of Hamiltonian \eqref{eq:ham} we have made use of the \textit{worm algorithm} (WA). The algorithm was originally introduced by Prokof’ev, Svistunov and Tupitsyn \cite{Prokofev1998} for quantum systems on a lattice, and then applied to classical spin models \cite{PhysRevLett.87.160601} too. Later the WA has been successfully extended also to off-lattice boson models \cite{PhysRevLett.96.070601,Cinti2010b,ciardi2024,PhysRevA.105.L011301}.

In its classical formulation, WA essentially gets rid of the critical slowing down problem and, at the same time, shows dynamical exponents which are comparable with those of the Wolff algorithm \cite{PhysRevLett.62.361}. It should be noted that, 
unlike the usual cluster methods, WA is based on a completely different description of the scrutinised system, featuring closed paths configurations and winding numbers, which are far from the standard \textit{site representation} \cite{9780511614460}. In this representation, the partition function can be defined as 
\begin{equation}\label{partfunc0}
Z = \sum_{\text{CP}} W_{\text{path}}\, ,
\end{equation}
where the sum in Eq.~\eqref{partfunc0} is taken over the whole space of closed path (CP) configurations, while $W_{\text{path}}$ is the weight of a specific closed path configuration. The expectation values for a generic observable $O$ (e.g., helicity modulus and energy per spin) likewise reads as
\begin{equation}\label{observable1}
\mean{O} = \sum_{CP} O_{\text{path}} \, W_{\text{path}} \,.
\end{equation}
where $O_{\text{path}}$ in Eq.~\eqref{observable1} is the value of the observable over the specified CP configuration. 
To perform a proper ``importance sampling'' \cite{9780511614460}, one generates states according to the distribution $p_{\text{path}}=W_{\text{path}}/Z$, choosing $W_{\text{path}}=e^{-\beta E_{\text{path}}}$, $\beta=1/T$, and $E_{\text{path}}$ being the energy associated to each CP configuration. 

Now we are going to explicitly show how to implement the WA for the spin system appearing in Hamiltonian \eqref{eq:ham}. For the sake of simplicity we start by looking at a simple two layers setup: $M=2$. Then the partition function reads as 
\begin{equation}\label{partfunc}
\begin{split}
    Z  &= \prod_i \int \frac{d \theta_i}{2 \pi} \frac{d \phi_i}{2 \pi} \, e^{- \beta H} \\
    & = \prod_i \int \frac{d \theta_i}{2 \pi} \frac{d \phi_i}{2 \pi} \, 
     \prod_{\langle i,j \rangle} e^{\beta J \cos(\theta_i - \theta_j)} \\
    &\times \prod_{{\langle i,j \rangle}} e^{\beta J \cos(\phi_i - \phi_j)} \prod_i e^{\beta K \cos(\theta_i - \phi_i)} \,,
\end{split}
\end{equation}
$\theta_i$ and $\phi_i$ being the spins on the first ($m=1$) and on the second layer ($m=2$).

\begin{figure}[t!]
\begin{center}
\includegraphics[scale=1.0]{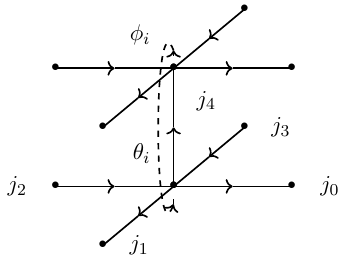}
\caption{\label{figbounds} Possible choice for bond orientations on the bilayer XY model lattice. The bond number is considered to be positive if the motion on the bond goes from left to right and back to front on each layer, and from bottom to top between the layers.}
\end{center}
\end{figure}

Accordingly with previous works \cite{PhysRevLett.87.160601,Nguyen2021,PhysRevB.100.064525}, we rewrite 
Eq.~\eqref{partfunc} over nearest neighbors as a product over bonds, but in the present case we must pay attention to separate bonds of the first layer, $b_1$, and those of the second, $b_2$. Ultimately, the last product of the partition function involves bonds between the two layers $b_z$. 
Furthermore, we Fourier expand the exponentials of the cosines:
\begin{eqnarray}
e^{\beta J\cos(\theta_i - \theta_j)} &=& \sum_{n_{b_1}}  I_{n_{b_1}}(\beta J) e^{i n_{b_1}(\theta_i - \theta_j)} \label{bessel1} \, ,\\
e^{\beta J\cos(\phi_i - \phi_j) }&=& \sum_{n_{b_2}} I_{n_{b_2}}(\beta J) e^{i n_{b_2}(\phi_i - \phi_j)}  \label{bessel2} \, ,\\
e^{\beta K\cos(\theta_i - \phi_i)}& =& \sum_{n_{b_z}} I_{n_{b_z}}(\beta K) e^{i n_{b_z}(\theta_i - \phi_i)} \label{bessel3} \, ,
\end{eqnarray}
where $I_{n_{b_1}}(\beta J)$, $I_{n_{b_2}}(\beta J)$, and $I_{n_{b_z}}(\beta K)$ are modified Bessel functions of the first kind \cite{Riley2006}. Substituting Eqs.~\eqref{bessel1}, \eqref{bessel2} and \eqref{bessel3} in Eq.~\eqref{partfunc} and switching the products over bonds with the respective sum over bond numbers, we finally arrive to the following equation:
\begin{equation}\label{partfunc1}
\begin{split}
    Z  &= \sum_{\{n_{b_1}\}} \sum_{\{n_{b_2}\}} \sum_{\{n_{b_z}\}}  \prod_{b_1} I_{n_{b_1}}(\beta J_1) \prod_{b_2} I_{n_{b_2}}(\beta J_2) \prod_{b_z} I_{n_{b_z}} (\beta K) \\ &\times  \prod_i \int \frac{d \theta_i}{2 \pi} \frac{d \phi_i}{2 \pi} \prod_{\langle i,j \rangle} e^{i n_{b_1}(\theta_i - \theta_j)} \prod_{\langle i,j \rangle} e^{i n_{b_2}(\phi_i - \phi_j)} \\  &\times \prod_{i} e^{i n_{b_z}(\theta_i - \phi_i)}\,.
\end{split}
\end{equation}
The integrals over the angular variables can be computed through bonds on the whole lattice. A configuration then is described by simple set of integer numbers, $n_{b_1}$, $n_{b_2}$ and $n_{b_z}$,  which are associated to specific links. A typical example is reported in Fig.~\ref{figbounds}. As an illustrative example, the figure shows a system where periodic boundary conditions along $z$ have been applied only \cite{tesimasini}. We can think about the set of integers as the number of times one passes over a link while moving along a path on the lattice. Usually these numbers are taken positive if the overall motion on the bond goes from left to right or from top to bottom on the link, otherwise negative.

To evaluate Eq.~\eqref{partfunc1}, we first focus on a single site $i$ whose contribution to the partition function yields:
\begin{equation}
\label{bound1}
\begin{split}
     & \int \frac{d \theta_i}{2 \pi}  \frac{\phi_i}{2 \pi} \, \prod_{k = 0}^{3} e^{i n_k (\theta_i - \theta_k)} \prod_{k=0}^{3}  e^{i m_k (\phi_i - \phi_k)} \,  e^{i n_4 (\theta_i - \phi_i)} \\
    & \propto  \int  \frac{d  \theta_i}{2 \pi}   e^{i \theta_i (n_0 + n_1 - n_2 - n_3 + n_4)}  \int \frac{d \phi_i}{2 \pi} e^{i \phi_i (m_0 + m_1 - m_2 - m_3 - n_4)}  \\
    & = \delta_{\, n_0 + n_1 +n_4; \, n_2 + n_3} \, \delta_{\, m_0 + m_1 ; m_2 + m_3 + n_4}  \\
    & = \delta^{(1)}_{\, n_{\text{in}}; \, n_{\text{out}}} \, \delta^{(2)}_{\, n_{\text{in}}; \, n_{\text{out}}}\,.
\end{split}
\end{equation}
One easily realises that the partition function results in a product of delta functions $\delta_{\, n_{\text{in}}, \, n_{\text{out}}} $, one for each of the $2\times L_x\times L_y$  spins (hereafter we take $L_x = L_y =L$). As a final outcome, the partition function 
can be expressed as a sum over the CP configurations
\begin{equation}
    Z = \sum_{\text{CP}} W(\{n_b\}, \beta)\, ,
\end{equation}
where the weight of a single configuration now reads
\begin{equation}
    W(\{n_b\}, \beta) = \prod_{b_1} I_{n_{b_1}}(\beta J) \prod_{b_2} I_{n_{b_2}}(\beta J) \prod_{b_z} I_{n_{b_z}}(\beta K) \,.
    \label{eq:weitos}
\end{equation}
The functional form of the weights $\eqref{eq:weitos}$ results adaptable without any difficulty to a multilayer geometry, as we can rewrite it, whatever the values of $M$, as:
\begin{equation}
    W(\{n_b\}, \beta) = \prod_{b} I_{n_{b}}(\beta{\mathcal J}_b) \,,
    \label{eq:weitM}
\end{equation}
taking into account that the argument of the Bessel function only changes depending on the interaction ${\mathcal J}_b$ on the bond $b$, i.e. ${\mathcal J}_b=J$ if the bond connects sites on the same layer, or ${\mathcal J}_b=K$ for inter-layer bonds.

Regarding the evaluation of the relevant physical estimators (namely energy and helicity modulus), also in this case we have to operate expressing these observable in terms of sums over CP. Looking at the total energy we get:
\begin{equation}
\label{totalenergy}
    E = - \frac{1}{Z} \frac{\partial Z}{\partial \beta} = \bigg \langle \sum_{b} J_{b} \frac{I^{\prime}_{n_b}(\beta {\mathcal J}_b)}{I_{n_b}(\beta {\mathcal J}_b)} \bigg \rangle \, ,
\end{equation}
$I^{\prime}_{n_b}$ being the first derivative of the modified Bessel functions,
while $\langle\cdots\rangle$ is a statistical average over the sampled bonds.

The critical properties of the system in Eq.~\eqref{eq:ham} 
can be scrutinized by using the generalized elasticity \cite{chaikin2000principles} (i.e. the response of the system to a small distortion) that can be defined by looking at the dependence of the free energy $F$ on a proper small twist $\mathbf{\Delta} = (\Delta_x, \Delta_y, \Delta_z)$ applied between the boundary spins of the sample. The global minimum of $F(T, \Delta)$ then corresponds to the limit of zero twist $\Delta\to0$. Expanding the free energy around this minimum one obtains 
\begin{equation}
\begin{split}
F(T,\Delta) &- F(T,0) =   \left. \frac{\partial^2 F(T, \Delta)}{\partial \Delta^2} \right|_{\Delta = 0} \frac{\Delta^2}{2!} \\ 
&+ \left. \frac{\partial^4 F(T, \Delta)}{\partial \Delta^4} \right|_{\Delta = 0} \frac{\Delta^4}{4!}  	
+ \cdots \,,
\label{eq:sviluppino}
\end{split}
\end{equation}
as odd powers of $\Delta$ can not appear in the expansion, being the sign of $\Delta$ irrelevant for the variation of $F$ induced by the twist . 
The coefficient of the second order term in the expansion~\eqref{eq:sviluppino} defines the 
helicity modulus. The latter 
is 
a non-negative function of $T$ and is going to dominate for small $\Delta$ \cite{chaikin2000principles}. 

As shown in the following, the helicity modulus can be written in terms of \textit{winding numbers} \cite{poll87}. 
For the sake of clarity we recall the procedure considering just one layer connected to the set of bounds $\{n_{b_1}\}$; we will see as the extension to a bilayer or even multilayer geometry is indeed straightforward. An easy way to impose a twist to a single slab (uncoupled to the others) is to apply the transformation
\begin{equation}
\left( \theta_i - \theta_j \right) \to \left( \theta_i - \theta_j -\frac{\mathbf{r}_{ij} \cdot \mathbf{\Delta}}{L} \right) \, ,
\end{equation} 
and taking $\mathbf{\Delta} = (\Delta_x, \Delta_y,0)$, whereas $\mathbf{r}_{ij}$ is $\hat{\mathbf{x}}$ or $\hat{\mathbf{y}}$. The additional term implies an alteration of the partition function that affects the Fourier expansion \eqref{bessel1} only:
\begin{equation}
\label{fourierchanging}
\begin{split}
    e^{\beta J \cos\left(\theta_i - \theta_j - \frac{\mathbf{r}_{ij} \cdot \mathbf{\Delta}}{L} \right) } & = \\ &\sum_{n_{b_1}} \left[ I_{n_{b_1}}(\beta J) \, e^{-i n_{b_1} \frac{\mathbf{r}_{ij} \cdot \mathbf{\Delta}}{L}} \right] e^{i n_{b_1} (\theta_i - \theta_j) }\, .
    \end{split}
\end{equation}
Analogous expressions hold for Eq.\eqref{bessel2} and \eqref{bessel3}. At the same time,  $W_1(\{n_{b_1}\}, \beta)$ changes as  
\begin{equation}
\label{twistedweight}
    W_1(\{n_{b_1}\}, \beta, \Delta) = \prod_{b_{1}} I_{n_{b_1}}(\beta J) e^{-i n_{b_1} \frac{\mathbf{r}_{b_1} \cdot \mathbf{\Delta}}{L}} \, ,
\end{equation}
where the subscript $\mathbf{r}_{ij}$ moves to $\mathbf{r}_{b_1}$, consistently with our description in terms of links. Notice that the terms containing the angular variable remain unaltered with respect to the torsion-less case. Thus, when computing the partition function, we still obtain the delta functions $\delta_{\, n_{\text{in}}; \, n_{\text{out}}}$ which control the CP configurations.

Taking advantage of the partition function for the single slab $Z_{slab}(\Delta)$, 
we obtain: 
\begin{equation}
\begin{split}
&\Upsilon_{slab}(T)  = \left. \frac{\partial^2 F(T, \Delta)}{\partial \Delta^2} \right|_{\Delta = 0} \\
&= -\frac{T}{Z_{slab}(\Delta=0)} \left. \frac{\partial^2 Z_{slab}(\Delta)}{\partial \Delta^2} \right|_{\Delta = 0}  \\
& = -\frac{T}{Z_{slab}(\Delta=0)} \sum_{\text{CP}} \prod_{b_1} I_{n_{b_1}}(\beta J) \, \left[ - \frac{i}{L} \sum_{b_1} n_{b_1} (\mathbf{r}_{b_1} \cdot \hat{\mathbf{n}}) \right]^2  \\
& = T  \bigg \langle  \frac{1}{L^2} \bigg[\sum_{b_1} n_{b_1} (\mathbf{r}_{b_1} \cdot \hat{\mathbf{n}})\bigg]^2  \bigg \rangle \, .
\label{eq:contaz}
\end{split}
\end{equation}
The quantity we are averaging is the square of the winding number associated to a path on the lattice, 
computed with respect to the direction of the twist $\hat{\mathbf{n}} = \hat{\mathbf{x}}$ (or $\hat{\mathbf{y}}$):
\begin{equation}
    \mathcal{W}_{\hat{\mathbf{n}}} = \frac{1}{L} \sum_{b_1} n_{b_1} (\mathbf{r}_{b_1} \cdot \hat{\mathbf{n}}) \, .
    \label{eq:windingo}
\end{equation}

The extension of Eq.~\eqref{eq:contaz} to a multi-layer systems ($M\ge2$) can be carried out exploiting the linearity of the task of torsion along the principal axes of the structure as described in Eq.~\eqref{eq:ham}. This means that one can defines a {\it parallel} helicity modulus, $\Upsilon^\parallel_{tot}$, which can be sampled as the square of a linear combination of the winding numbers belonging to each slab, that is, to each set of bonds $\{n_{b_1}\}$, $\{n_{b_2}\}$,\ldots$,\{n_{b_M}\}$, respectively. By choosing the same twist for both layers, the parallel stiffness reads as 
\begin{equation}\label{upsilontot}
\begin{split}
\Upsilon^{\parallel}_{tot} (T) & =  T \Big \langle \frac{1}{L^2} [ \sum_{b} n_b ]^2 \Big \rangle \\ & = T \Big \langle \frac{1}{L^2} \bigg[ \sum_{b_1} n_{b_1} (\mathbf{r}_{b_1} \cdot \hat{\mathbf{n}}) + \sum_{b_2}n_{b_2}(\mathbf{r}_{b_2} \cdot \hat{\mathbf{n}})  \bigg]^2 \Big \rangle \\
&= T \Big \langle (\mathcal{W}_{1,\hat{\mathbf{n}}} + \mathcal{W}_{2,\hat{\mathbf{n}}} +\ldots+   \mathcal{W}_{M,\hat{\mathbf{n}}}  )^2  \Big \rangle \, ,
\end{split}
\end{equation}
where $\mathcal{W}_{1,\hat{\mathbf{n}}}$, $\mathcal{W}_{2,\hat{\mathbf{n}}}$,\ldots, $\mathcal{W}_{M,\hat{\mathbf{n}}}$ are the winding numbers of an $M$-layered systems. 

For the sake of completeness, a  {\it transverse} helicity modulus can be easily dug out as well.
We compute it by applying the twist in a direction orthogonal to the layers. So setting $\hat{\mathbf{n}} = \hat{\mathbf{z}}$ one gets
\begin{equation}
	\Upsilon^{\perp}_{tot} = T \Big \langle \frac{1}{L^2} \bigg[ \sum_{b} n_b \bigg]^2 \Big \rangle = T \langle \mathcal{W}_z^2 \rangle \, .
	\label{eq:ups_perp}
\end{equation}
By all means, the Section~\ref{section4} presents results only concerning the longitudinal helicity moduli, that is single slab $\Upsilon_{slab}$, Eq.~\eqref{eq:contaz}, and total-parallel $\Upsilon^\parallel_{tot}$, Eq.~\eqref{upsilontot}.


The third term of the expansion~\eqref{eq:sviluppino} introduces a fourth-order modulus, usually denoted as $\Upsilon_4$. Again we give a general expression for the bilayer along $\hat{\mathbf{x}}$ ($\hat{\mathbf{y}}$), i.e. $\Upsilon^{\parallel}_4$ and $\hat{\mathbf{z}}$, i.e $\Upsilon^{\perp}_4$ . Either way, we start off from Eq.~\eqref{eq:sviluppino} as follows 
\begin{equation}
\begin{split}
\Upsilon_4 &= \left. \frac{\partial^4 F(T, \Delta)}{\partial \Delta^4} \right|_{\Delta = 0} \\
& \\
 &= T \left. \left[ \frac{3}{Z^2} \left( \frac{\partial^2 Z}{\partial \Delta^2} \right)^2 - \frac{1}{Z} \frac{\partial^4 Z}{\partial \Delta^4} \right] \right|_{\Delta = 0}\,
\end{split}
\label{upsilon4}
\end{equation}
The first term of the second line in Eq.~\eqref{upsilon4} brings us straight to the helicity modulus. Working out on the last one, we have
\begin{equation}
\begin{split}
\left. \frac{\partial^4 Z}{\partial \Delta^4} \right|_{\Delta = 0} & = \sum_{\text{CP}} \left. \left[ \prod_b I_{n_b}(\beta J_b) \frac{\partial^4 }{\partial \Delta^4} \prod_b e^{-i n_b \frac{(\mathbf{r}_c \cdot \hat{\mathbf{n}})}{L} \Delta}  \right] \right|_{\Delta = 0} \\
& = Z \Big \langle \frac{1}{L^4} \bigg[ \sum_{b} n_b (\mathbf{r}_b \cdot \hat{\mathbf{n}} ) \bigg]^4 \Big \rangle \,,
\end{split}			
\end{equation}
so that the general formula for the fourth-order modulus along $M$-layers ($\hat{\mathbf{n}}=\hat{\mathbf{x}}$, $\hat{\mathbf{y}}$) reads:
\begin{equation}
\label{eq:ups4_par}
\Upsilon_4^\parallel  =  \frac{3}{T^2} (\Upsilon^\parallel)^2 - T  \Big \langle (\mathcal{W}_{1,\hat{\mathbf{n}}} + \mathcal{W}_{2,\hat{\mathbf{n}}} +\ldots+   \mathcal{W}_{M,\hat{\mathbf{n}}} )^4  \Big \rangle \, .
\end{equation}
As pointed out 
in \cite{PhysRevB.67.172509} for the single-layer case, the helicity modulus has the usual 
jump to zero at $T_{BKT}$ only if the fourth-order modulus approaches a nonzero and negative value around $T_{BKT}$. 


\section{Results}\label{section4}

\begin{figure}[t!]

\centering

\includegraphics[scale=0.40]{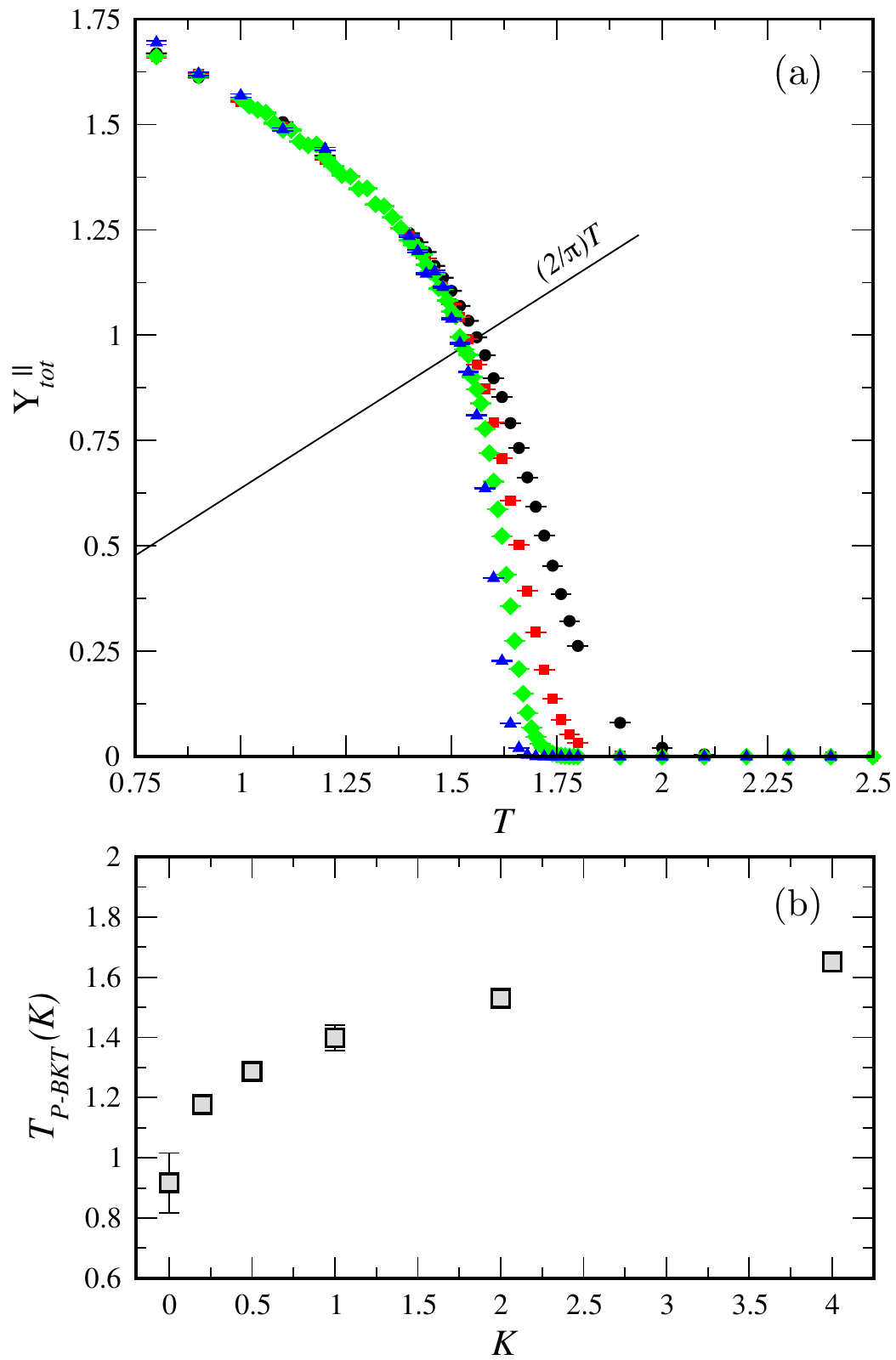}

\caption{(a) Helicity modulus~\eqref{upsilontot} versus temperature for the system with $K=2$ and size $L$ = 16 (black circles), 32 (red squares), 64 (green diamonds) and 128 (blue triangles). Straight line represents $2T/\pi$.(b) $T_{P-BKT}(K)$ as a function of the parameter $K$ (see main text).}

\label{fig_Y_due}

\end{figure}

In this Section we aim to present the results of Monte Carlo simulations of a bilayer, which have been carried out on Hamiltonian~\eqref{eq:ham} by using the classical WA method for multilayer systems discussed in Section~\ref{section1bis}. 

Monte Carlo simulations have been performed for a number of layers $M=\,$1 and 2, and the properties of the system have been investigated as the inter-layer coupling changes, by considering $K$=0.2, 0.5, 1, 2, 4.  Finite-size scaling was studied taking $L$ between 16 and 128; the statistical errors of the simulations were evaluated making use of the usual blocking technique \cite{newman1999monte}.

We first illustrate a general feature of the bilayer structure such as the total helicity modulus, evaluated when the twist is applied along the $\hat{\mathbf{x}}$ or $\hat{\mathbf{y}}$ direction. Figure \ref{fig_Y_due}(a) depicts $\Upsilon^{\parallel}_{tot}$, see Eq.~\eqref{upsilontot}, as a function of the temperature $T$ for the lattice sizes $L=16,32,64,128$. 

For the sake of clarity, in Fig.~\ref{fig_Y_due}(a) we present the case for $K=2$. For $T\to 0$, $\Upsilon^{\parallel}_{tot}$ is going to be equal to $K$. As can be observed, $\Upsilon^{\parallel}_{tot}$ follows the universal scaling behavior expected for a two-dimensional-like system. 
In order to have an estimate of the critical temperature of the bilayer system with respect to a twist acting on both layers, we assume that the universal jump 
is $2/\pi$, according to the classical result by Nelson and Kosterlitz \cite{nel77} for the BKT transition. 
Looking at the intersection of the black line representing $(2/\pi)T$ with the simulation data for $\Upsilon^{\parallel}_{tot}(T)$,  we can extrapolate from Fig.~\ref{fig_Y_due}(a)
the value of a critical temperature, that we denote by $T_{P-BKT}(K)$, 
to refer that we are mentioning to the response of the bilayer system as a whole. To determine $T_{P-BKT}(K)$ 
in the thermodynamic limit we apply the following relation: 
\begin{equation}
\label{tbkt}
T_{P-BKT}(K,L)=T_{P-BKT}(K)+\frac{c_1}{[\ln(c_2L)]^2}\,,
\end{equation}
where $c_1$ and $c_2$ are constants, while $T_{P-BKT}(K,L)$ 
is the estimate of the transition temperature for the system having size $L$ \cite{PhysRevB.65.184405}. By using the fitting relation~\eqref{tbkt}, we get $T_{P-BKT}(K=2)=1.52(3)$; 
the same analysis has been done for all the other considered values of $K$, to get $T_{P-BKT}(K)$. 

The same analysis proposed in Fig.~\ref{fig_Y_due}(a) and in Eq.~\eqref{tbkt} has been carried out also for the other values of the inter-layer coupling constant. Figure~\ref{fig_Y_due}(b) depicts $T_{P-BKT}(K)$ as a function of $K$. By increasing the inter-layer coupling the system enlarges its rigidity along the perpendicular direction to the planes. Breaking this rigidity requires overcoming stronger thermal fluctuations, which, as observed in Fig.~\ref{fig_Y_due}(b), leads to an increase in $T_{P-BKT}$. Therefore, it seems reasonable to assume that the vortex coupling between layers drives 
a BKT critical behavior of the entire system.

\begin{figure}
\begin{center}
\includegraphics[scale=0.4]{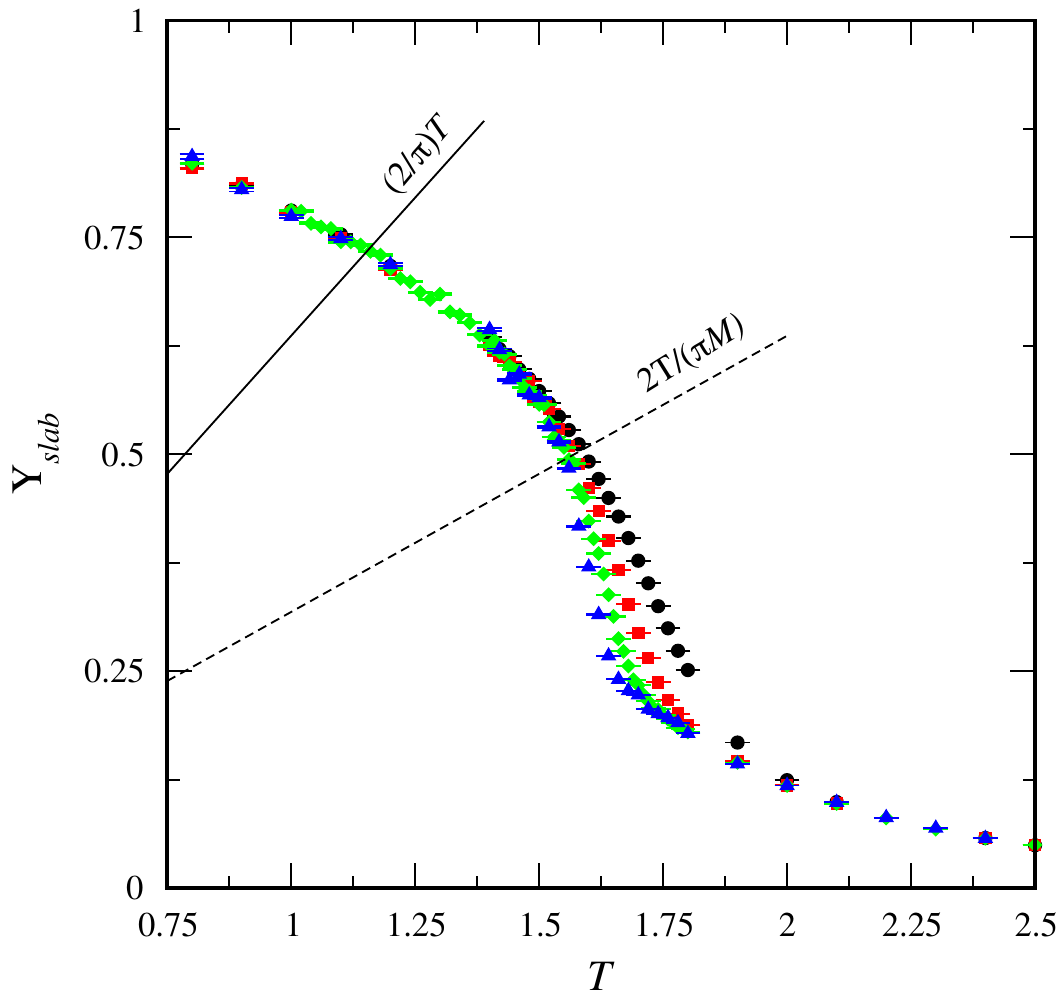}
\caption{$\Upsilon_{slab}(T)$ versus temperature for $K=2$. $L$ = 16 (black circles), 32 (red squares), 64 (green diamonds) and 128 (blue triangles). Dashed line representing the line $2T/(\pi M)$, $M=2$, whereas the straight one $2T/\pi$.}
\label{fig_Y}
\end{center}
\end{figure}

Figure~\ref{fig_Y} shows the estimator of the single slab helicity modulus $\Upsilon_{slab}$, Eq.~\eqref{eq:contaz}, for different $L$ and $K=2$. It is apparent that this quantity does not depend on the system-size from $T\approx2.0$ onward.
$T_{P-BKT}(2)$ seems to be compatible with the the crossing point between $\Upsilon_{slab}(T,L)$ and the dashed line shown in Fig.~\ref{fig_Y}. Interestingly, the line is not referring to the intersection point  $\Upsilon_{slab}(T)/T=2/\pi$ but $\Upsilon_{slab}(T)/T=2/(\pi M)$ (here $M=2$). In the latter case $\Upsilon_{slab}(T)$ displays a ``quasi-discontinuous'' behavior which seems to be coherent with the analysis we provided for $\Upsilon^{\parallel}_{tot}$ and $T_{P-BKT}(2)$. To wrap up, also taking into account what was discussed above regarding Fig.~\ref{fig_Y_due}(b), we can reasonably infer that $T_{P-BKT}(K)$ signalises the onset of a quasi-long-range order featuring a vortex anti-vortex coupling between the two layers (BKT-paired regime).

\begin{figure}[t!]
\begin{center}
\includegraphics[scale=0.27]{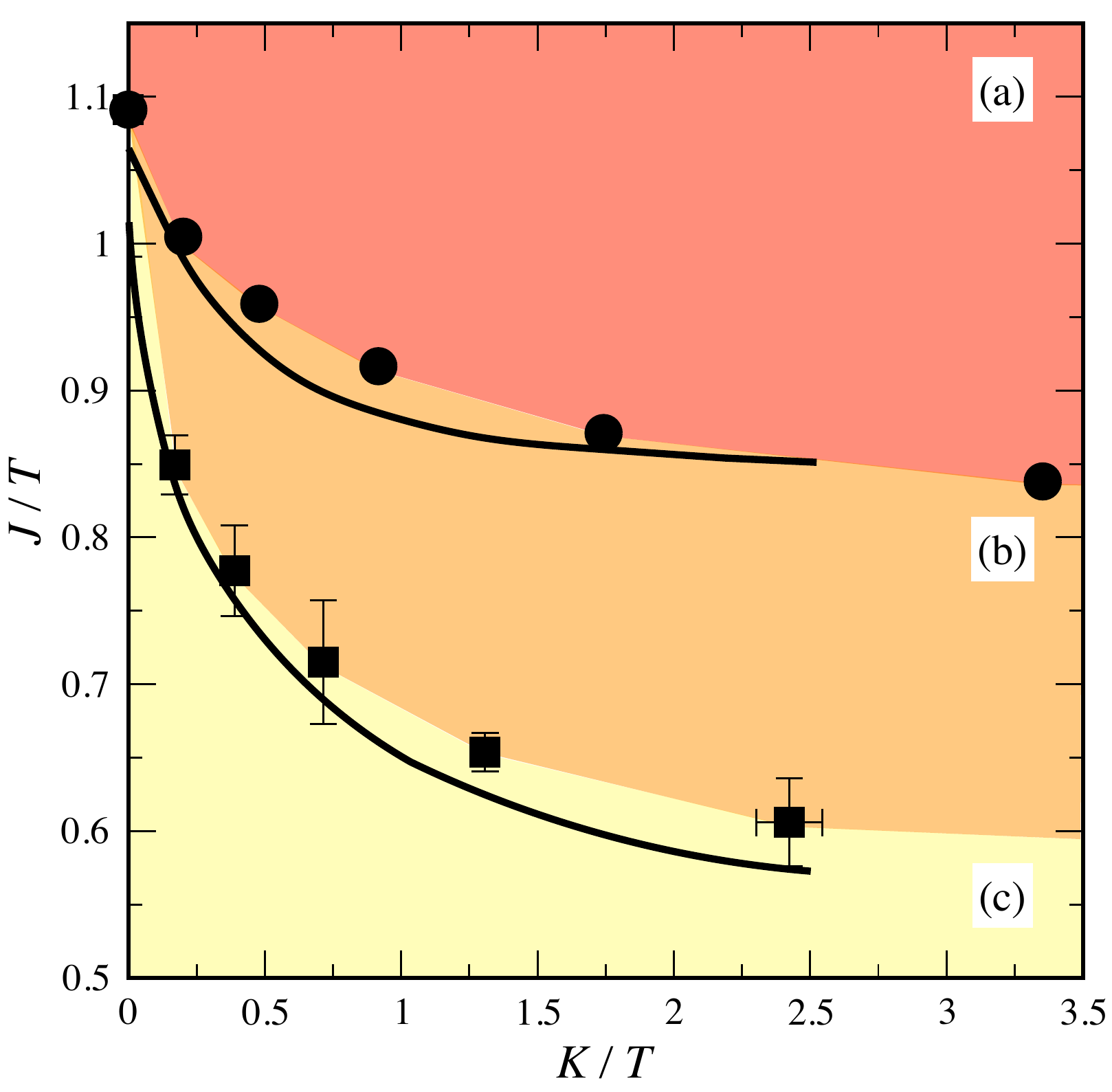}
\caption{Phase diagram of $J/T$ versus $K/T$. (a) Phase characterized by an algebraic decay of correlations along all directions; (b) BKT-paired phase; (c) disordered phase. Circles denote $T_{NK}(K)$ whereas squares report $T_{P-BKT}(K)$, see main text. Solid lines are extracted from the data in Fig.1 of Ref.~\cite{PhysRevLett.123.100601}, which were obtained by considering the behavior of the correlation function at different temperature and $K$ for $L=64$ and up to $K/T=2.5$ .} 
\label{phasediag}
\end{center}
\end{figure}

$\Upsilon_{slab}(T)/T=2/\pi$ is instead found at a lower temperature with respect to $T_{P-BKT}(2)$, that is $T_0\approx1.5$ (continuous line in Fig.~\ref{fig_Y}). According to Ref.~\cite{nel77}, at $T\lesssim T_0$ and $L\to\infty$, the single layer possesses a finite helicity modulus. This corresponds to the presence of an in-layer quasi-long-range order. We denote the temperature of this second interception point with $T_{NK}(K)$.
It is worthwhile stressing that for $T< T_{NK}(K)$ the system described by Eq.~\eqref{eq:ham} inherently holds an inter-layer quasi-long-range order. Therefore, it turns out that for $ T_{NK}(K) < T < T_{P-BKT}(K)$ the bilayer will maintain a robust BKT-paired regime, where the appearing of the inter-layer correlations display an algebraic behavior as discussed in Ref.~\cite{PhysRevLett.123.100601}. Strikingly, in this windows of temperatures, $\Upsilon_{slab}(T)$ shows no dependence on $L$, thus excluding the presence of a universal jump at $T_{NK}$ and for $L\to\infty$, of $\Upsilon_{slab}(T)$
defined from the response of the system to a torsion acting on the separated layers. This might be regarded as an indirect estimation of the BKT-paired phase boundary, albeit we point out that the extrapolation of $T_{NK}(K)$ results consistent with the two-point correlation function findings shown in Ref.~\cite{PhysRevLett.123.100601} (see the following discussion on Fig.~\ref{phasediag} for further details). Finally, we note that for $K \to 0$, we must have $T_{NK}\equiv T_{BKT}$, consistently with an XY single layer, in which $T_{BKT} \approx 0.893$ \cite{gupta1988,komura2012}.


\begin{figure}[t!]
\begin{center}
\includegraphics[scale=0.39]{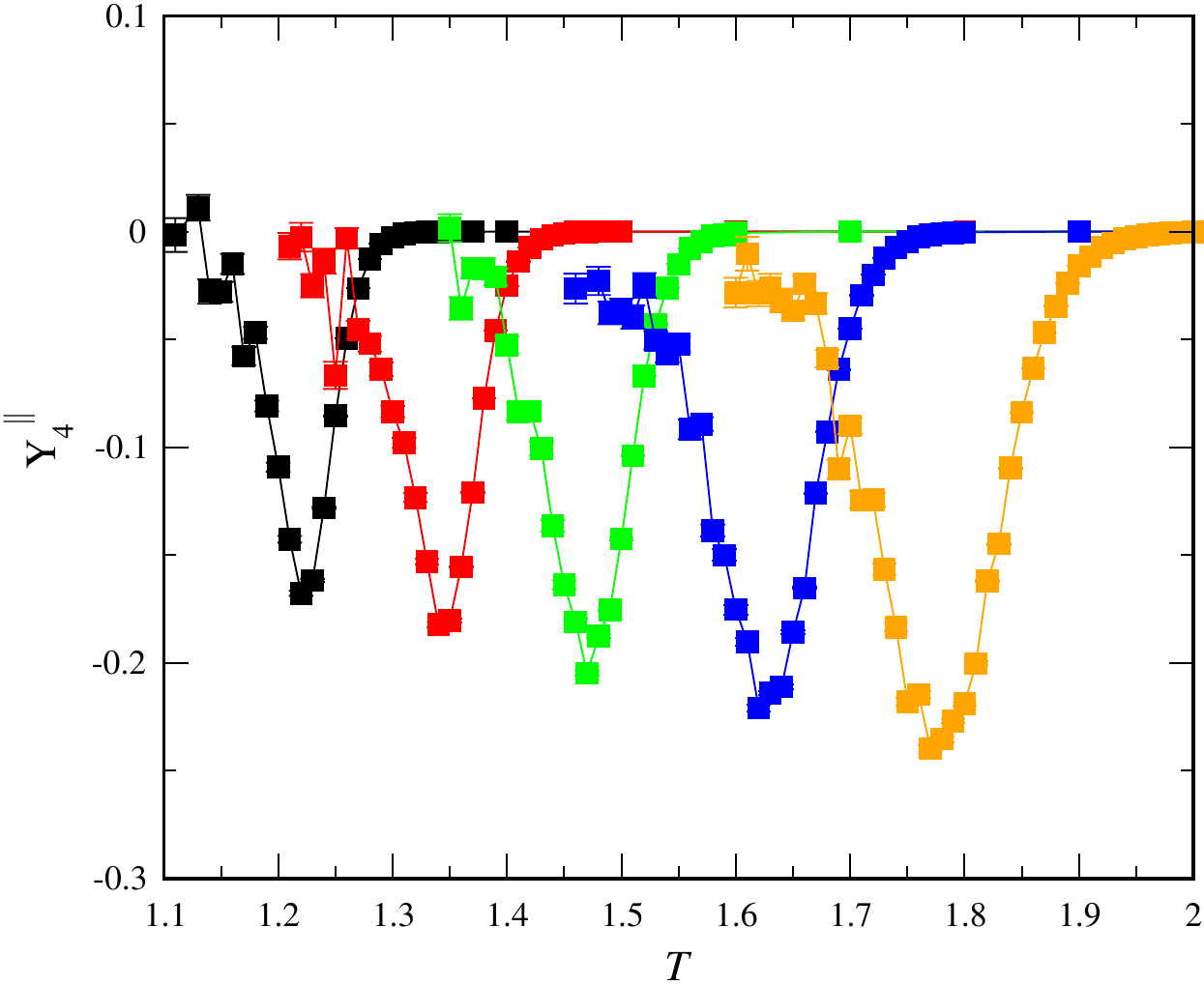}
\caption{Parallel fourth order modulus for the bilayer XY model versus the temperature at different values of the interlayer coupling $K$ = 0.2 (black), 0.5 (red), 1.0 (green), 2.0 (blue), and 4.0 (orange).}
\label{fig_y4}
\end{center}
\end{figure}

The character of the BKT regime at $T_{P-BKT}(K)$ can be verified by considering the parallel fourth order modulus for the system as obtained in Eq.~\eqref{eq:ups4_par} and depicted in Fig.~\ref{fig_y4}. The figure presents simulations for $L=64$ for different values of $K$. As a general consideration, $\Upsilon_4^\parallel$ goes to zero at high temperature and it shows a peculiar dip at temperatures immediately above $T_{P-BKT}$, that is, corresponding to the region where the helicity modulus drops to zero. Regarding the bilayer, such features seem to be confirmed in Fig.~\ref{fig_y4}, where these valleys are connected to the 
transition related to $\Upsilon^{\parallel}_{tot} (T)$. In general the position of the minimum of the fourth order modulus moves towards lower temperatures as $L$ increases (not shown here). It has been pointed out that for the single layer ($K=0$) the width and depth of the dip decreases as $L$ increases \cite{PhysRevB.67.172509}. In particular, for $L\to\infty$ the depth reaches a nonzero value. 

The comparison of $T_{P-BKT}(K)$ and the temperature signalising the in-layer algebraic correlations, $T_{NK}(K)$, is reported in Fig.~\ref{phasediag}, where we present a phase diagram of $J/T$ as a function of $K/T$. Here $T_{NK}(K)$  (circles) and $T_{P-BKT}(K)$ (squares) results are clearly separated. Such a split is even present at small $K$, whereas by increasing $K$ the separation gets more and more robust (see orange area). As a consequence, it seems fair to associate the $T_{P-BKT}$ to the quasi-long-range order between layers, regardless the in-layer one.
Figure~\ref{phasediag} also includes a comparison with the analysis of correlation functions at different temperatures and inter-layer couplings (see black solid lines) done in Ref.~\cite{PhysRevLett.123.100601}. It is worth noting that the distinction between algebraic and exponential behavior of the function was there made by using a purely empirical criteria, rather than through a finite-size scaling analysis. Starting from the lower line in Fig.~\ref{phasediag}, we observe a surprisingly good agreement with $T_{P-BKT}(K)$ obtained in Eq.~\eqref{tbkt}, considering also the fact that the data are obtained using two significantly different Monte Carlo methods. 

On this matter, the worm algorithm proves to be 
more flexible and easier to use compared to the complexities of the probability-changing cluster algorithm \cite{PhysRevB.65.184405} in investigating the transition regions. By definition, during the sampling procedure the worm is able to reach any spin-spin link regardless their belonging to one plane or the other, and without imposing any conditions at all (see Section~\ref{section1bis}). As a result, this methodology can furnish a complete spectrum of the observables allowing to probe the critical properties related to the free energy~\eqref{eq:sviluppino}. 

The upper part of the phase diagram (red region, $T\lesssim T_{NK}(K)$) instead identifies the phase featuring algebraic correlations along all directions. 
The upper line in Fig.~\ref{phasediag} represents the border that separates exponential from power-law behavior of the in-layer correlations as obtained in Ref.~\cite{PhysRevLett.123.100601}. The comparison between the results obtained by the correlation function fitting in Ref.~\cite{PhysRevLett.123.100601} and from the analysis carried out on $\Upsilon_{slab}(T)$ appears to be good overall. However, as previously observed, the evaluation of $T_{NK}(K)$ does not rely on finite-size scaling argument as in the obtained results for the single slab helicity modulus no apparent size dependence is observed. 

  Finally, the lower part of Fig.~\ref{phasediag} (yellow region) concerns the disordered phase of the model. It is worthwhile stressing that the phase diagram agrees with Ref.~\cite{PhysRevLett.123.100601} where the systems has been investigated by sampling the in-layer and out-of-layer correlations. 

\section{
Conclusions}\label{section5}

In this work we investigated the 
properties of Hamiltonian~\eqref{eq:ham} by adapting a classical worm algorithm capable of sampling bilayer (or even multilayer) geometries with anisotropic interactions and spins featuring an $O(2)$ continuum symmetry. The implemented methodology has proven to be flexible and stable, especially in sampling helicity modulus and fourth-order modulus along any needed direction. 
It turns out that, to gain a comprehensive understanding of the critical properties of Hamiltonian~\eqref{eq:ham}, it is essential to evaluate these statistical observables: 
indeed, they have provided novel and meaningful insights into the nature of the two critical temperatures shown in Fig.~\ref{phasediag}. We defined and studied a total helicity, when the twist is acting on the whole system \cite{fisher1973}, and a slab helicity, when the twist acts on the separate layers. We showed that this distinction proves to be crucial to characterize the superfluid properties of the bilayer systems in its phases.

By decreasing $T$, and for any inter-layer coupling considered, we get first a transition temperature, $T_{P-BKT}(K)$, signalizing the onset of a BKT-like phase.
The analysis of $\Upsilon^{\parallel}_{tot}(T)$ appear to indicate that the system establish a quasi-long-range-order between layers (BKT-paired phase). Then, we spotted a second temperature, $T_{NK}(K)$, which is given at the intersection of $\Upsilon_{slab}$ and the line $2T/\pi$. We can associate $T_{NK}(K)$ with the temperature that marks the establishment of an algebraic decay of in-layer correlations.

The findings discussed above look like to bears out reasonably well with Ref.~\cite{PhysRevLett.123.100601}. In this paper authors provided a similar phase diagram with respect to the one 
shown in Fig.~\ref{phasediag} by looking at the trend of the correlation functions as well as employing renormalization group methods. Therefore, by evaluating measurable physical quantities directly connected to the BKT transition [including free energy too, see Eq.~\eqref{eq:sviluppino}], our investigation results are especially relevant since they quantitatively confirm the existence of a BKT-paired phase. 

This preliminary study needs further insights to better characterize the paired phase both for larger system sizes and with more layers ($M>2$), possibly in presence of different couplings in different layers, and also to provide the necessary support for its experimental verification. For example, feasible experimental platforms could be found in ultracold atoms \cite{PhysRevA.95.023622}, molecular magnets on surfaces \cite{Briganti2023,Lunghi2015}, ultra-thin films~\cite{PhysRevB.79.134420} and semiconductor systems \cite{PhysRevX.10.031007}. We stress the point that $T_{P-BKT}$ has been located by using as value for the jump of the total helicity modulus $\Upsilon^\parallel_{tot}$ the same universal value of the (single-layer) 
BKT universality class \cite{nel77}, a point which certainly deserves a further investigation. Similarly, a full characterization of both transition lines is a very interesting point to be further clarified. Finally, we mention it would be very interesting to compare the helicity modulus for $M$ increasing in the multi-layer model with the corresponding results for the 3D layered model.


\begin{acknowledgments}
We acknowledge L. Salasnich for many useful discussions in the initial stage of the work presented in this paper. Discussions with G. Bighin, N. Defenu, R. Konik and G. A. Williams are also gratefully acknowledged.
We thank the High Performance Computing (CHPC) in Cape Town for providing computational resources (research programme Grant No. PHYS0892). This work was also supported by the European Union through the Next Generation EU funds through the Italian MUR National Recovery and Resilience Plan, Mission 4 Component 2 - Investment 1.4 - National Center for HPC, Big Data and Quantum Computing (CUP B83C22002830001). A. C., A. T. and F. C. acknowledge financial support from PNRR MUR Project No. PE0000023-NQSTI. 
A.T. acknowledge support for the CNR/MTA Italy-Hungary
2023-2025 Joint Project "Effects of strong correlations in interacting many-body systems and quantum circuits".
\end{acknowledgments}


%

\end{document}